 \documentclass[useAMS,usenatbib]{mn2e}

\newcommand{\swift}{{\it Swift }}
\newcommand{\rxte}{{\it RXTE }}
\usepackage{graphicx,times}
\begin{document}
\title{A failed outburst of H 1743-322}
\author[Capitanio et al. ] {F.~Capitanio$^{1}$, T. Belloni$^{2}$, M. Del Santo$^{1}$, P. Ubertini$^{1}$ \\
 $1$ INAF IASF-Roma, Via Fosso del Cavaliere 100, 00033 Rome,Italy \\
$2$ INAF, Osservatorio Astronomico di Brera--- Italy \\ }
\def\LaTeX{L\kern-.36em\raise.3ex\hbox{a}\kern-.15em
    T\kern-.1667em\lower.7ex\hbox{E}\kern-.125emX}
\newtheorem{theorem}{Theorem}[section]
\date{Released 2002 Xxxxx XX}

\maketitle
 \begin{abstract}
We report on a campaign of X-ray and soft $\gamma$-ray observations of the black hole candidate H~1743-322 (also named IGR J17464-3213), performed with the {\it RXTE}, INTEGRAL and \swift satellites. The source was observed during a short outburst between 2008 October 03 and 2008 November 16. The evolution of the hardness-intensity diagram  throughout the outburst is peculiar, in that it does not follow the canonical pattern through all the spectral states (the so called {\it q-track} pattern) seen during the outburst of black-hole transients. 
On the contrary, the source only makes a transition from the Hard State to the Hard-Intermediate State. After this transition, the source  decreases in luminosity and its spectrum hardens again. This behaviour is confirmed both by spectral and timing analysis. 
 This kind of outburst has been rarely observed before in a transient black hole candidate

\end{abstract}

\section{Introduction}



During the outbursts of transient black hole candidates (BHCs), in addition to the large changes in X-ray luminosity, marked variations are observed in the properties of the timing and the energy spectrum 
often on very short time scales (see e.g. \citealt{belloni2005}). 
We still do not have a detailed understanding of all the mechanisms that lead to changes in the X-ray emission properties, but the physics involves the structure of the accretion flow around the black hole as well as the connection between the accretion disc and the steady or impulsive jets that can be emitted from these systems.  The main cause of the changes in the X-ray emission properties is the variation of the mass accretion rate onto the black hole; however some phenomena indicate that other parameters are also important (e.g. \citealt{Homan01}).

From the observational point of view, the emission properties of accreting black holes are often classified in terms of observed spectral and timing parameters. From their combination, a number of source states have been identified (\citealt{belloni2005,belloni2009}, for an alternative definition, see \citealt{remmc}). 

The high-energy spectra can be described as the combination of a soft thermal component together with a hard power law component. The latter component often shows a cutoff at high energies~\citep{Tanaka}. This decomposition is the simplest phenomenological model. However for the hard component complex models can be used as for example the Comptonization models (see e. g. \citealt{Tit, PP}){\it . The} simplest decomposition is less model  dependent and provides at least a qualitative measurement of the behaviour of the source.
The power density spectra, used to characterise the fast variability, show a combination of power-law and Lorentzian components. When the Lorentzians are zero-centered, they are referred to as ``band-limited noise'', while if they are narrow they are called {\it Quasi-Periodic Oscillations} (QPO, see e.g. \citealt{bpk}). Another important observable that can be used to trace source states is the total amount of variability in the 0.1-64 Hz band, expressed in terms of fractional integrated rms (see e.g. \citealt{belloni2005,hombel,belloni2009}).

\begin{figure}
\centering
\includegraphics[angle=0, scale=0.5]{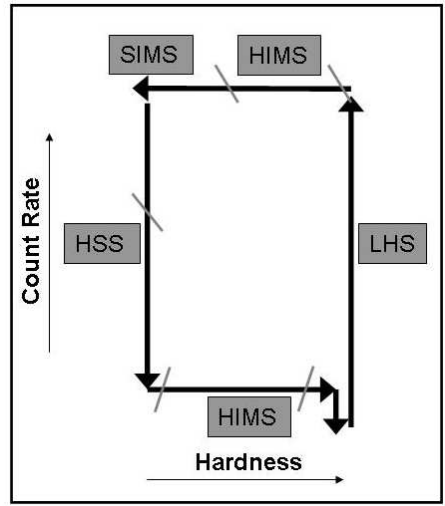}
\caption{Sketch of the "canonical q-track pattern" through all the different spectral states (for more details see ~\citealt{hombel} and ~\citealt{belloni2009}).}
\label{figtesi}
\end{figure}

A BHC spends its time mostly in a quiescent state at low flux level  ($<$ 10$^{32}$--10$^{33}$ ergs s$^{-1}$,  e.g. \citealt{Camp}). When the outburst begins, the luminosity of the source increases and the X-ray spectrum is dominated by a power law component with a hard photon index  of $\sim$ 1.4-1.5 and a high-energy cutoff around 100 keV (low/hard state, hereafter LHS). The radio emission in this state indicates the presence of steady jets, while the power spectrum is dominated by a strong band limited noise ($>$30\% fractional rms). 
Then the outburst  evolves as the source increases its luminosity and its spectrum starts to change: the soft thermal component appears and becomes increasingly important, the energy peak of the emission softens and the photon index of the hard component steepens ($\sim$ 2.0-2.5). Two different states with these spectral characteristics have been defined: the hard intermediate state (HIMS) and the soft intermediate state (SIMS). The characteristics of these two states are quite complex: the changes can be established mostly by the timing properties \citep{hombel} and also by the ejection of relativistic jets associated to the transition from HIMS to SIMS.
After the SIMS, the source enters a state where the X-ray spectrum is dominated by the emission of the soft thermal component (high/soft state, hereafter HSS). A non-thermal power law tail is also present without any detectable cutoff, while the power spectrum is characterised by a low-level (1-2\% fractional rms) variability.
Then the flux starts to decrease, most likely following a parallel decrease in accretion rate. At some point, a reverse transition is started and the path is followed backwards all the way to the LHS and then to quiescence. As mentioned above, the luminosity level of this back-transition is always lower than that of the corresponding forward-transition.
The description above is the basic general pattern (the so-called "{\it q-track}" pattern in the hardness-intensity diagram (see Figure~\ref{figtesi}); several examples of "{\it q-track}" patterns are reported in \citealt{hombel} or in Dunn et al. 2009 in prep.), which has been modeled after repeated outbursts of GX 339-4 \citep{fbg2004,belloni2005,hombel,mela, belloni2009}. During the HSS, minor transitions to SIMS and HIMS have been observed also in GX 339-4 (e.g. \citealt{Pg,belloni2005,mela2}). Other sources behave in a more complicated way, but the general classification into four states holds for all of them (see e.g. \citealt{mike2000,frontera2001,chaty2003,hynes2003}). 
Interestingly, until now all black-hole transients have shown two types of behaviour: after the initial LHS, most sources show a transition to the HIMS at a luminosity level which is always different and might be related to the previous history of the transient \citep{yu2007}.  If this transition takes place, the source always reached the HSS. However, a few sources (both NS and BHC X-ray binaries) never left the LHS at all as reported for example by \citet{broc} for V404 Cyg, A1524-62, 4U1543-475, GRO J0422+32, GRO J1719-24, GRS1737-21 and GS 1354-64, by \citet{Stu} for XTE1550-564 and by \citet{Rod} for Aql X-1. The only possible exception to this dichotomy is represented by SAX~J1711.6--38 (see \citealt{wm02}), a faint transient X-ray binary classified as BHC~\citep{Liu}.

 In this paper we present the results of the {\it RXTE}, \swift and INTEGRAL data analysis of the last outburst of the recurrent transient H~1743--322 during which only two states were sampled: the LHS and the HIMS.  It is the first time that this kind of outburst is analysed in  detail showing that the evolution of the BHC outbursts, through the different spectral states, is still difficult to predict.

\subsection{Short History of H~1743--322}

On 2003 March 21 {\it INTEGRAL} (MJD=52719) detected  a relatively bright source ($\sim$60 mCrab at 15-40 keV) named IGR J17464--3213. The source was localised at R.A.\ (2000) $=17^{\rm h}46.3^{\rm m}$, Dec.\ $=-32\degr 
 14.4^\prime$, with an error box of 1.6 arcmin (90\% confidence) and then  associated with H~1743--322 \citep{mark,rev}, a bright BHC observed by {\it HEAO-1} in 1977 with an intensity of 700 mCrab in  the 2-10 keV~\citep{dox}. The outburst evolution of 2003 was followed by \rxte and {\it INTEGRAL} reporting strong  flux and spectral variability \citep{parm,homan,cap2005,joinet}. 

H~1743--322, after the first and brightest outburst, also underwent two fainter outbursts on September 2004 and September 2005~\citep{atel301,atel575,cap2006}.
On January 2008 a third outburst was detected by {\it RXTE}/ASM~\citep{kal}. Then a \swift ToO was granted in order to follow the source evolution~\citep{CapATel2,CapATel1}. 
Seven months later (on September 23, MJD=54732), another outburst was detected by INTEGRAL during the  Galactic bulge monitoring~\citep{kuul} showing that the source was  in a hard state with an increasing flux. {\it Swift}, \rxte and INTEGRAL followed the outburst evolution. 
 Furthermore, on October 23 (MJD=54762) an \rxte observation of H~1743--322  indicated that the source had undergone a state transition from the LHS to the HIMS~\citep{belATel}. The study reported in this paper is focused on the last part of this outburst.  Some preliminary results of these observations were already reported in various ATels  (see e.g. \citealt{Ricci} and reference therein), while results on the early phase of the outburst are presented in \citet{pr}.

\begin{figure}
\begin{center}
 \includegraphics[angle=90, scale=0.37]{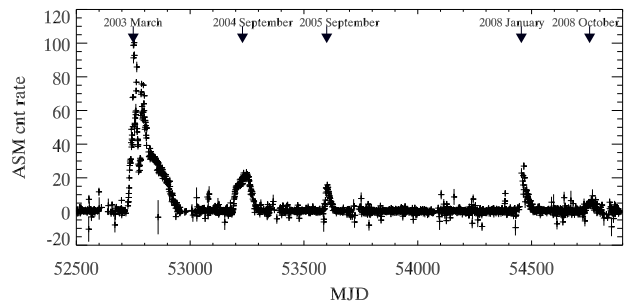}
      \caption{ASM 1.5-12 keV  one-day averaged light curve of  H~1743--322 from 2003 until the end of 2008.}
         \label{asmtot}
\end{center}
 \end{figure}

\section{Observation and analysis}

The \rxte campaign of pointed observations covers the period of the outburst starting from MJD=54740 (2008 October 10) to MJD=54775 (2009 November 5) for a total of 37 pointings with a total exposure time of 65.5 Ks. The PCA and HEXTE data analysis was performed with the standard \rxte software within {\tt HEASOFT V6.6}  following the standard extraction procedure. For spectral analysis only PCA2 for PCA and cluster B for HEXTE were used. A systematic error of  0.6\% ~\citep{Wilms} was added to the PCA spectra. For the fitting, the energy ranges 3-20 keV and 20-130 keV were used for PCA and HEXTE respectively. 

For the timing analysis of the PCA data, for each of the observations, we produced power spectra from 64-s stretches accumulated in the channel band 0-35 (2-15 keV) with a time resolution of 1/1024 s. The resulting power spectra were then averaged, resulting in one power spectrum per observation. The power spectra were normalised according to the description  by \citealt{Leahy} and converted 
to squared fractional rms~\citep{BelHas,Miy}. The contribution due to Poissonian statistics was subtracted (see \citealt{Zhang}). The timing analysis was performed with custom software. 


During the 2008 October outburst  a public INTEGRAL ToO campaign was carried out (\citealt{Ricci} and references therein).
Three observations were performed respectively in 2008 October 10  (65 ks, MJD=54749), October 22 (86 ks,MJD=54761) and October 28 (80 ks,MJD=54767
) with a total of 136 science windows (SCW)\footnote{ A SCW is a unit of INTEGRAL continuous observing time that is at the base of the INTEGRAL data processing.} of about  2500 s each  (INTEGRAL revolutions 732, 734 and 736).   In order to obtain a wider energy range for the spectra, our analysis was focused on ISGRI~\citep{Lebr}, the low energy detector of the $\gamma$-ray  telescope IBIS~\citep{ube}.  The data of the INTEGRAL spectrometer SPI were not used because of the low angular resolution of the instrument (for details see \citealt{Ver}). For the ISGRI data analysis we used the latest release of the standard Offline Scientific Analysis (OSA) version 7. The ISGRI energy range considered for the fitting is 20-200 keV with a systematic error of 2\%   as usual in the INTEGRAL spectral analysis (see also \citealt{Jourdain}). 
The ISGRI light curve was obtained by extracting the source count rate from the images  in the 40-100 keV band for each SCW.

A \swift ToO was also performed  with three pointings. We extracted the XRT spectra of the Window-Timing (WT) mode pointings ( less affected by pile-up) in order to better constrain the equivalent hydrogen column. The data were processed with the standard \swift tools: 
XRT software version 0.12.2 and FTOOLS version 6.6.2 
Only the second WT mode {\it Swift}/XRT observation was performed simultaneously with \rxte and INTEGRAL and thus it was used to obtain a joint spectrum with the other instruments (see  the bottom panel of Figure~\ref{counts2}).
The \swift/BAT transient monitor light curve was provided by the \swift/BAT team\footnote{http://swift.gsfc.nasa.gov/docs/swift/results/transients/index.html}.

\section{Results}

\subsection{Outburst Evolution}
\begin{figure}
\centering
\includegraphics[angle=90, scale=0.34]{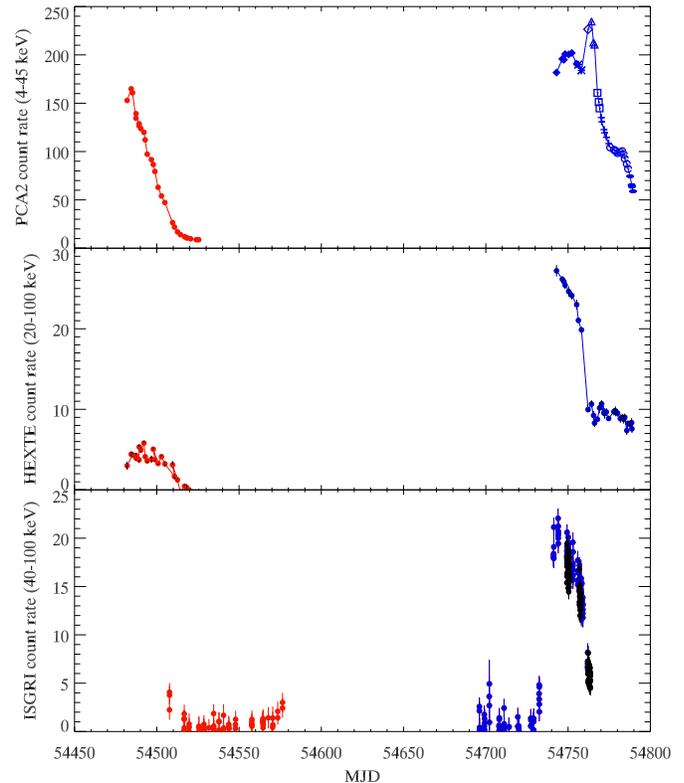}
\caption{ Light curves of both 2008 January and October H~1743--322 outbursts in different energy ranges: in red the October outburst, in blue the January one. Top panel: \rxte/PCA 4-45 keV light curve (for the different symbols of the blue curve see Figure~\ref{licu}). Middle panel:  \rxte/HEXTE 20-100 keV light curve. Bottom panel: IBIS/ISGRI 40-100 keV light curve binned to a SCW; the data are taken from the {\it INTEGRAL Galactic Bulge Monitoring}. The black circles represent the three ToO observations analysed in this paper.}
   \label{bat_vs_asm}
 \end{figure}

\begin{figure}
\centering
\hspace{-1.4cm}
\includegraphics[angle=90, scale=0.35]{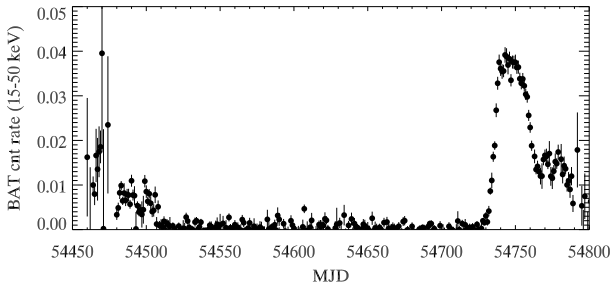}
\caption{H1743-322 BAT daily averaged light curve, 15-50 keV}
   \label{bat_mon}
 \end{figure}

 The 2008 October outburst of H 1743-322 was quite short, and fainter  than the previous ones (see Figure~\ref{asmtot}).  In Figure~\ref{bat_vs_asm} we show the temporal behaviour of H 1743-322 in different energy ranges. 
 For a direct comparison with the previous outbursts of the  source, we show in the figure, as an example, also the light curves of the January 2008 outburst. During the January 2008 outburst the source reached the HSS~\citep{kal} showing the standard {\it q-track} behaviour as for the previous outbursts of H1743-322 (see for example \citealt{cap2005,cap2006,Mc}).
The top panel and the middle panel represent respectively the 4-45 keV ({\it RXTE}/PCA) and the 20-100 keV ({\it RXTE}/HEXTE) light curves. Both curves are binned to a single observation. The bottom panel shows the INTEGRAL IBIS/ISGRI  40-100 keV light curve, binned to a SCW~\footnote{ The data are taken from both the public archive of the INTEGRAL Galactic Bulge Monitoring., http://isdc.unige.ch/Science/BULGE/ and from the 2008 public INTEGRAL ToO.}. 
The first outburst, was observed by \rxte only during the return to the quiescent state. The monitoring of the most recent outburst (blue curve) was more complete even though \rxte missed the rising phase of the outburst and its coverage ended before the full return to quiescence.
The INTEGRAL monitoring started at the end of the January outburst; a good coverage was only achieved during the October outburst (bottom panel of Figure~\ref{bat_vs_asm}). 
Figure~\ref{bat_mon} shows the \swift/BAT daily averaged light curve (15-50 keV) of the two 2008 outbursts of H1743-322.



\begin{figure}
   \centering
\hspace{-1.4cm}
\includegraphics[angle=90, scale=0.34]{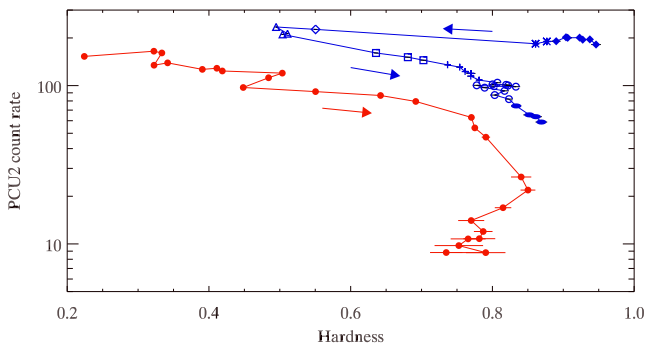}
      \caption{ \rxte/PCU2 hardness-intensity diagram  (hardness is defined as the ratio of the count rates in the 11-20 keV and 4-10 keV bands) of the two 2008 outbursts of H 1743-322. The different symbols of the blue curve are associated with their averaged spectrum of Figure~\ref{spectra} and Table~\ref{spec_res} as follow: filled diamonds for {\it (a)}, asterisks for {\it (b)}, open diamonds for {\it (c)}, triangles for {\it (d)}, squares for {\it (e)}, crosses for {\it (f)}, circles for {\it (g)} and ellipses for {\it (h)}.}
         \label{licu}
 \end{figure}


The  HIDs of the two outbursts are  compared in Figure~\ref{licu}: for the first outburst  (red curve), the PCA caught the source at the end of the {\it q-track} diagram (see \citealt{belloni2005,belloni2009}), when the energy spectrum of H~1743--322 was hardening again through the HIMS, then moving vertically downward along the LHS branch returning to quiescence. 
For the second outburst, after the initial LHS rise (missed by \rxte) the source moves horizontally to the left, softening, then jumps to a much softer location, from which it slowly returns to the hard track along a diagonal path. Clearly, the final hard state is also missed. As Figure~\ref{licu} shows, the softest points of the second outburst reach only  intermediate values of the hardness ($\sim$0.5) that correspond in the previous outburst to the HIMS.
 This fact is confirmed also by the HEXTE light curve of the two outbursts (see middle panel of Figure~\ref{bat_vs_asm} red and blue curves): the ratio between the PCA and HEXTE fluxes shows that the October outburst is clearly harder than the January one.

\subsection{Timing Analysis}
\begin{figure}
   \centering
 \includegraphics[angle=0, scale=0.41]{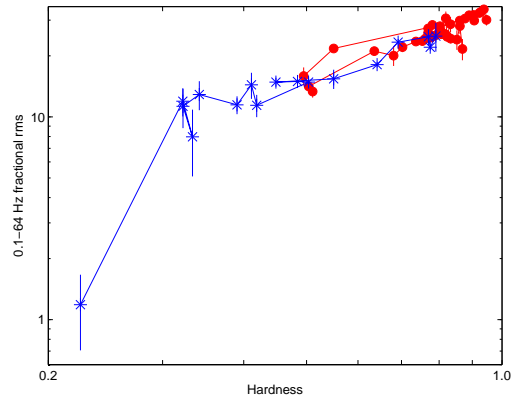}
      \caption{Hardness-rms diagram of the two 2008 outbursts of H 1743-322. The blue asterisks indicate the January outburst, red circles the October one. Note that 1.7 days is the average distance between two subsequent observations (maximum gap 4.5 days) }
         \label{fig:rms}
 \end{figure}

Figure \ref{fig:rms} shows the Hardness-rms diagram for both 2008 outbursts. Also from this figure it is evident that the October outburst saw the source remaining at a high level of variability, with integrated fractional rms  always above 10\%. In contrast, in January the sampling of the final part of the outburst, started at a low hardness and little variability (around 1\%).
Together with the HID, this figure suggests that in October the source never left the HIMS. 

Inspection of the power-density spectra of the October outburst confirms this hypothesis. Band-limited noise is seen in all cases. All observations show a type-C QPO  (see ~\citealt{rem02}) with the exception of that of October 25 (MJD=54764), the softest of the sample. However, the observations before and after, on October 23 (MJD=54762) and 27 (MJD=54766), show a QPO evolving from 5.6 Hz to 6.7 Hz, with an rms decreasing from 5.5\% to 3.1\%. The 3$\sigma$ upper limit for a 6 Hz QPO with the same FWHM (around 0.4 Hz) for October 25 is 3.3\%, which makes the non-detection compatible with neighboring observations.

From timing analysis, we can conclude that all observations of the October outburst indicate H 1743-322 being in the HIMS, a state characterised by the presence of a type-C QPO and intermediate hardness.

\subsection{Spectral Analysis}
\begin{figure}
   \centering
 \includegraphics[angle=90, scale=0.35]{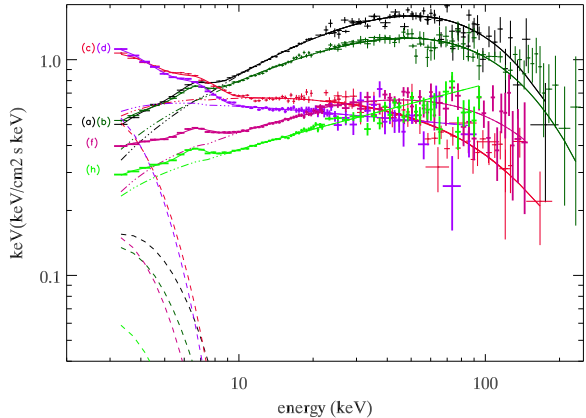}
      \caption{\rxte/PCU and  INTEGRAL IBIS/ISGRI spectral evolution of H 1743-322 during the 2008 October outburst: see Table~\ref{spec_res} for the spectral parameters value of each spectrum. See Figure~\ref{licu} for the position of each spectrum in the HID.}
         \label{spectra}
 \end{figure}

We analysed all the \rxte pointed observations in order to study the source spectral behaviour. The data were fitted with a simple model consisting  of an absorbed disc blackbody plus a cutoff power-law component. From the analysis of the Swift/XRT data an equivalent hydrogen column value of  N$_{H}$=(1.6$\pm$0.1)$\times$10$^{22}$ atoms cm$^{-2}$ was derived  and in all the other fits the N$_{H}$ was fixed to the value derived from the \swift data analysis. 
To account for cross-calibration problems between the three different instruments (PCA, HEXTE, IBIS/ISGRI), multiplicative constants were added to the fits. An emission line with centroid fixed at 6.4 keV was needed to obtain  good fits.
 The relative change of the disc inner radius is derived from the square root of the disk black body component normalisation constant (for details see ~\citealt{bb}). 
After a detailed study of each pointing, we averaged the spectra of contiguous observations with consistent spectral parameters. Table~\ref{spec_res} summarises our results, while the unfolded spectra of different groups of observations are presented in Figure~\ref{spectra}. Each spectrum was rebinned with {\tt HEASOFT V6.4} tool {\tt GRPPHA} in order to get an adequate signal to noise ratio.
Concerning the first group of spectra, {\it (a)} in Table~\ref{spec_res}, the best fit model is described by a disk black body with an internal temperature of about 1 keV plus a high-energy power law component with a photon index of about 1.3 and a cutoff of 75 keV. The second group of spectra,  {\it (b)}, is characterised by a softening of the photon index together with the decrease of the  0.1-500 keV flux (see Table~\ref{spec_res}). 


Between the observations {\it(b)} and {\it (c)}, the spectrum changes fast: the cutoff reaches a value of about 100 keV  (red curve in Figure~\ref{spectra}). Two days after (spectrum {\it (d)})  the cutoff is no longer detectable; at the same time the photon index becomes softer.
The disc black body inner radius increases its value  by about 70\%  during the softening  while the inner temperature remains contant (see Table~\ref{spec_res}).

\begin{figure}
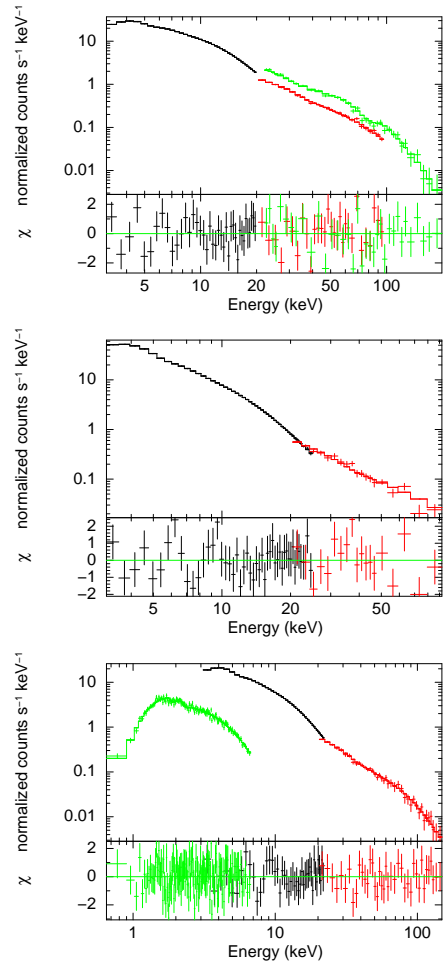

   \centering
 \includegraphics[angle=-90, scale=0.22]{ldatadel1.ps}
\includegraphics[angle=-90, scale=0.22]{ldatadel_d.ps}
\includegraphics[angle=-90, scale=0.22]{ldatadel_swift_new.ps}

\caption{Spectra {\it (a)} (Top panel), {\it (d)} (middle panel) and {\it (f)} (bottom panel) plotted in counts (see  Table~\ref{spec_res} for spectral parameters).}
\label{counts2}
\end{figure}

After October 30 the source spectra {\it (e)} and {\it (f)}  harden again:  in accordance with the HID, the cutoff is again detectable  at about 109 keV. Both the inner radius and the photon index approach values similar to those observed in the first two groups of observations (see Table~\ref{spec_res}).
 We show in Figure~\ref{counts2} (bottom panel) the (f) spectrum fitted jointly with the simultaneous {\it Swift}/XRT window timing observations. This spectrum confirms the presence of the disk black body component and the fit parameters are all consistent within the errors with the RXTE spectrum (see Table~\ref{spec_res}). While the top and the middle panels show respectively the {\it (a)} and {\it (d)} spectra.

At the end of the outburst (spectra {\it (g)} and {\it (h)}) the flux slightly continues to decrease, the inner temperature of the disk still remains unvaried,  while it is no more possible to constrain the cutoff.  Note, however that the last INTEGRAL observation, that permits to extend the spectrum up to 200 keV, was performed in the 2008 October 23 (MJD=54762) (spectrum (c)). Thus,  the spectra taken after  October 23 cover only energy range from 3 to 130 keV. This fact limits the possibility to constrain the cutoff  for observations after that date.

\begin{table*}
\begin{minipage}{220mm}
\begin{tabular}{lccccccccccc}
\hline
ID    & date    & date             & T$_{in}$               & R$^{*}_{in}$   &$\Gamma$               & E$_{c}$         & FLUX$_{(0.1-500)}$          & FLUX$_{(2-10)}$ & FLUX$_{(10-100)}$  & $\chi^{2}_{red.}$  & d.o.f. \\
\hline
  -    &    mm/dd  &  MJD  & keV  & Km & -  & keV  & (erg s$^{-1}$cm$^{-2}$) & { \small (erg s$^{-1}$cm$^{-2}$)}& {\small (erg s$^{-1}$cm$^{-2}$)}   & -  &- \\
  -      &   (2008)   &  -               & -                      & -              & -                      &  -            &   $\times$10$^{-9}$ & $\times$ 10$^{-9}$ & $\times$ 10$^{-9}$ &-&-\\
\hline
{\it (a)}& 10/03-10/16 &{\bf 54742-54755} & 1.1$^{+0.1}_{-0.1}$    &4 $^{+3}_{-1}$  &1.23$^{+0.03}_{-0.03}$ & 64$^{+3}_{-3}$  & 8.9    &1.7   &5.0        & 1.1                & 96 \\  
{\it (b)}&10/17-10/19  &{\bf 54756-54758} & 1.0$^{+0.1}_{-0.1}$    &5$^{+2}_{-1}$   &1.45$^{+0.03}_{-0.03}$ &84$^{+6}_{-6}$   &8.1     &1.7  &4.1        &1.0                 & 104\\     
{\it (c)}&10/23 &{\bf 54762}       & 0.83$^{+0.03}_{-0.03}$ &16$^{+2}_{-1}$  &1.9$^{+0.1}_{-0.1}$    &109$^{+27}_{-19}$&10    &2.8    &2.0        & 1.1                & 92\\         
{\it (d)}&10/25-10/27&54764-54767       & 0.79$^{+0.03}_{-0.02}$ & 20$^{+2}_{-2}$ & 2.10$^{+0.03}_{-0.03}$& --              & 12    &2.9    &2.1        &1.0                 & 61\\              
{\it (e)}&10/28-10/30 &54767-54769       & 0.72$^{+0.04}_{-0.04}$ &15$^{+4}_{-2}$  &1.95$^{+0.03}_{-0.03}$ & --              & 7.1   & 1.8   &2.2        &1.0                & 65\\                 
{\it (f)\footnote{ {\it Swift}/XRT, {\it RXTE}/PCA and {\it RXTE}/HEXTE joint spectrum} } &10/31-11/04&54770-54774      & 0.82$^{+0.03}_{-0.03}$ &6$^{+4}_{-2}$   &1.60$^{+0.04}_{-0.05}$ &104$^{+16}_{-26}$& 5.1    &1.2   &2.1        &0.9                & 215\\     
{\it (g)}&11/07-11/16 &54776-54786       & 0.80$^{+0.1}_{-0.1}$ &5$^{+2}_{-1}$   &1.71$^{+0.02}_{-0.02}$ & --              &6.7     &0.9   &2.1      & 1.0                & 68\\                     
{\it (h)}&11/18-11/19 &54788-54789       & 1.0$^{+0.1}_{-0.1}$    & 3$^{+2}_{-1}$  &1.60$^{+0.03}_{-0.04}$ &--               & 5.3    &0.6   &1.7        & 1.1                & 51\\
 \hline
 \end{tabular}
\end{minipage}
\caption{Best fit spectral parameters of the seven groups of spectra. INTEGRAL and 
\rxte  commonly fitted spectra are represented by the MJD date in bolt.
 Note: all the errors are at 90\% confidence level. T$_{in}$: the inner temperature of the disk;
 R$^{*}_{in}$=(R$_{in}$/D$_{10}$)$\times\sqrt[]{cos(i)}$: 
where R$_{in}$ is the inner radius of the disk in km; {\it i} is the 
inclination angle of the system and D$_{10}$ is the distance of the 
source in unit of 10kpc; $\Gamma$: power law photon index; E$_{c}$: 
high energy cutoff; FLUX$_{(0.1-500)}$: unabsorbed  flux  extraploted
 from the models in the 0.1-500 keV energy range; Flux$_{(2-10)}$: 
unabsorbed flux between 2-10 keV; Flux$_{(10-100)}$: unabsorbed flux
 between 10-100 keV; $\chi^{2}_{red}$: reduced $\chi$ square.} 
\label{spec_res}
\end{table*}

\section{discussion} 

At first inspection the shape of the HID of H 1743-322 suggests a normal evolution of the October outburst: (missed) hard state, followed by a HIMS, a fast jump to the soft state (with or without a sampling of the SIMS), then a return path at lower flux. 
 However, the softest points are only at an intermediate hardness, which, in the previous outburst,  for example, corresponded to the HIMS. The hardness is only a rough indication of the spectral shape and similar states have been seen to correspond to slightly different hardness values even in different outbursts of the same source. Anyway the lack of soft states is an important fact and possibly suggests that  the soft state was never reached. This fact, only supposed by the study of the HID, is confirmed by the results of the timing analysis. 
 The spectral analysis, in accordance with the HID, sampled the softening of the source: the flux of the black body component increases and the   R$_{in}$ decreases. This means that the disk  approaches the last stable orbit, while, curiously, the T$_{in}$ remains substantially unvaried being also quite high for a HIMS.

As far as we know, this is the first time that an outburst, that  left the LHS and does not reach the HSS, has ever been studied in detail. Searching for similar cases in the literature, we have found only one outburst comparable with our results, the one of SAX J1711.6--3808 in 2001.  This outburst was not covered very densely, but its softest observation showed a power spectrum and a total fractional rms typical of the HIMS, while the slope of the hard part of the energy spectrum hardly reached 2.4~\citep{wm02}. Also in this case we have an HIMS with an  relatively high inner-disk temperature as 0.86$\pm$0.04 keV~\citep{wm02} and a luminosity consistent with the softest state of H 1743-322.
We conclude that we observed from H 1743-322  a failed outburst. In fact the source,  even  if softens, it never reaches the HSS  during the 2008 October outburst.

The range in luminosities of hard-to-soft transitions in black-hole binaries observed with \rxte is 0.2-1 L$_{Edd}$~\citep{Chen}. Interestingly, the lowest transitions, at 0.2 L$_{Edd}$, are those observed from Cyg X-1, which also does not reach very soft spectral hardnesses (see \citealt{Wilms,belloni2009}). Similarly, we found that the luminosity of the softest state  of H 1743-322 outburst  (spectrum (d) in Table~\ref{spec_res}), computed in unit of Eddington luminosity, is L$\sim$0.1 L$_{Edd}$ ( considering a mass of 10 M$_{\odot}$ and
a distance of 10  kpc as estimated for H 1743-322 by \citealt{Mc}). 
 Although the mass accretion rate is a very important parameter involved in the transition, another parameter seems to prime the transition out from the hard state. This second parameter, whose nature is still not clear~\citep{Es,Homan01}, drove the October 2008 transition from the LHS to HIMS.  Then  probably an accretion rate decrease did not permit to continue the canonical transition pattern to the HSS. This means that the source did not pass the jet-line in the HID (see \citealt{hombel}) and probably there was not any  jet major ejection ~\citep{fbg2004}. This despite the fact that the inner disk radius was seen to decrease  and according to the ~\citet{fbg2004} model the acceleration of the jet to higher Lorentz factor had already started.
The results reported in this paper demonstrate that the processes starting at the beginning of the outburst can be reversible even after the transition to HIMS.

  Other cases previously presented in literature as failed outbursts of transient X-ray binaries, are LHS-only outbursts  without any sign of state transitions  at all (see e.g. \citealt{broc} and \citealt{Stu}). Conversely, the data presented here show that the full pattern (LHS, HIMS, SIMS, HSS) and  LHS-only pattern are not the only two possibilities for the temporal evolution of a BHC outburst.

 The October 2008 outburst of H 1743-322, showing only LHS and HIMS, takes place at low luminosity (L$\sim$0.1 L$_{Edd}$) and the lack of soft-state transitions is probably connected to   a  premature decrease of the mass accretion rate.  Again, this bring us back to one of the major problems for the interpretation of the spectra/timing evolution of the outbursts: what physical parameter determines the  transitions starting from the low hard state.



         \section*{Acknowledgements}
     
       This work has been supported by the Italian Space Agency through grants  I/008/07/0 and I/088/06/0. 
       TMB acknowledges support from the International Space Science Institute.
       We acknowledge the use of public data from the \swift  data archive and all the \swift team for its support. Our particular thanks goes to Dr. J. M. Miller and his colleagues who immediately returned to the scientific community their INTEGRAL ToO data.

\end{document}